\documentclass[11pt,twoside]{article}
\usepackage{./asp2014}{}

\aspSuppressVolSlug
\resetcounters

\begin{document}

\title{Observations of magnetic fields in Herbig Ae/Be stars}

\author{S.~Hubrig$^1$, S.~P.~J\"arvinen$^1$, M.~Sch\"oller$^2$, T.~A.~Carroll$^1$, I.~Ilyin$^1$, M.~A.~Pogodin$^3$}
\affil{$^1$ Leibniz-Institut f\"ur Astrophysik Potsdam (AIP), An der Sternwarte~16, 14482~Potsdam, Germany}
\affil{$^2$ European Southern Observatory, Karl-Schwarzschild-Str.~2, 85748 Garching, Germany}
\affil{$^3$ Central Astronomical Observatory at Pulkovo of the Russian Academy of Science,
Pulkovskoye chaussee 65, 196140 Saint-Petersburg, Russia}

\paperauthor{S. Hubrig}{shubrig@aip.de}{}{Leibniz-Institute f\"ur Astrophysik Potsdam
(AIP)}{}{Potsdam}{}{14482}{Germany} 
\paperauthor{S.~P. J\"arvinen}{}{}{Leibniz-Institute f\"ur Astrophysik Potsdam
(AIP)}{}{Potsdam}{}{14482}{Germany}
\paperauthor{M. Sch\"oller}{}{}{European Southern
Observatory}{}{Garching}{}{85748}{Germany}
\paperauthor{T.~A. Carroll}{}{}{Leibniz-Institute f\"ur Astrophysik Potsdam
(AIP)}{}{Potsdam}{}{14482}{Germany}
\paperauthor{I. Ilyin}{}{}{Leibniz-Institute f\"ur Astrophysik Potsdam
(AIP)}{}{Potsdam}{}{14482}{Germany}
\paperauthor{M.A. Pogodin}{}{}{Central Astronomical Observatory at
Pulkovo of Russian Academy of Sciences}{}{Saint-Petersburg}{}{196140}{Russia}

\begin{abstract}
Models of magnetically driven accretion reproduce many observational properties 
of T\,Tauri stars. For the more massive Herbig Ae/Be stars, the corresponding picture 
has been questioned lately, in part driven by the fact that their magnetic fields  are 
typically one order of magnitude weaker. Indeed, the search for magnetic fields in 
Herbig Ae/Be stars has been quite time consuming, with a detection rate of about 7\%, 
also limited by the current potential to detect weak magnetic fields. Over the last two 
decades, magnetic fields were found in about twenty objects and for only two Herbig 
Ae/Be stars was the magnetic field geometry constrained. Further, studies were 
undertaken to investigate the time dependence of spectroscopic tracers of 
magnetospheric accretion. Overall, it seems that while there is proof that magneospheric accretion is 
present in some Herbig Ae stars, there is less evidence for the Herbig Be stars.
\end{abstract}

\section{Introduction}

Magnetic fields are found at all stages of stellar evolution, from young 
T\,Tauri stars to the end products of stellar evolution:
white dwarfs and neutron stars.
The incidence of magnetic fields in stars is diverse. The presence of a convective envelope 
is a necessary condition for significant magnetic activity in stars and magnetic activity is found 
all the way from late A-type stars (e.g.\ in Altair; \citealt{RobradeSchmitt2009}) 
with very shallow convective envelopes down 
to the coolest fully convective M-type stars. All solar type stars appear to be magnetic, the 
stronger the more rapidly they rotate (e.g.\ \citealt{Pallavicini1981}). 
This is understood through the $\alpha\Omega$--dynamo, 
which is thought to operate in the convective envelope of these stars. In intermediate mass 
main sequence stars, only about 10\% are found to have kG--strength large--scale magnetic fields. 
Here, the correlation of magnetic fields with stellar rotation is opposite to that of solar type stars. 
Whereas most intermediate mass stars are rapid rotators throughout their main sequence life, 
the magnetic stars are mostly slow rotators.

The question on the origin of magnetic fields in massive and intermediate mass stars with 
radiative envelopes is still unanswered and under debate. It has been argued that magnetic 
fields could be fossil relics of the fields that were present in the interstellar medium 
from which the stars have formed (e.g.\ \citealt{Moss2003}). However, the fossil field 
hypothesis has problems. One of them comes from studying the distribution of magnetic 
Ap stars in the H--R diagram. 
It was shown by \citet{Hubrig2000} that these stars are concentrated towards the center of 
the main-sequence band and almost no magnetic star can be found close to the zero age main sequence. No clear picture 
emerged in this work as to the possible evolution of the magnetic field strength across the main 
sequence. Alternatively, magnetic fields may be generated by strong binary interaction,
in stellar mergers, or during a mass transfer or common envelope evolution 
\citep{Tout2008}. The resulting strong 
differential rotation \citep{Petrovic2005} is considered as a key ingredient for the generation of magnetic fields.
The evolution of 
magnetic field configurations in Ap and Bp stars with masses between 1.6 and 5.7\,$M_{\odot}$ was 
considered until now only in a single study \citep{Hubrig2007}. 
No similar work has been carried out yet for more massive early B-type and O-type stars.

Studies of magnetic fields in stars at early evolutionary stages, before they arrive on the main sequence, 
are of special interest to get an insight into the magnetic field origin. It is generally accepted 
that magnetic fields are important ingredients of the star formation process 
(e.g.\ \citealt{McKeeOstriker2007}) and are 
already present in stars in the pre-main sequence (PMS) phase. However, it is not clear yet whether they 
persist until the main--sequence evolution. Current theories are not able to present a consistent 
scenario of how the magnetic fields in 
Herbig Ae/Be stars are generated and how these fields interact 
with the circumstellar environment, consisting of a combination of disk, wind, accretion, and jets. 
On the other hand, understanding the interaction between the central stars, their
magnetic fields, and their protoplanetary disks is crucial for reconstructing the Solar
System's history, and to account for the diversity of exo-planetary systems.

\section{Magnetic field studies}

The PMS T\,Tauri stars stand out by their strong emission in chromospheric and transition-region lines.
The presence of magnetic fields in higher mass PMS stars, 
the so-called Herbig Ae/Be stars, has long been suspected, in particular on the account of H$\alpha$ 
spectropolarimetric observations pointing out the possibility of the existence of a physical transition 
region in the H-R diagram from magnetospheric accretion, similar to that of classical T\,Tauri stars 
(e.g.\ \citealt{Vink2002}). 
While models of magnetically driven accretion and outflows successfully reproduce many observational 
properties of the classical T\,Tauri stars, the picture is completely unclear for 
the Herbig Ae/Be stars, due to the poor knowledge of their magnetic field topology. So far, 
the magnetic field geometry was constrained only for two Herbig Ae/Be stars, 
V380\,Ori \citep{Alecian2009} and HD\,101412 \citep{Hubrig2011}, 
and only about 20 Herbig stars were reported to host magnetic fields 
(\citealt{Hubrig2015} and references therein). 

The two Herbig Ae/Be stars exhibit a single-wave variation in the mean longitudinal 
magnetic field 
during the stellar rotation cycle. This behaviour is usually considered as evidence 
for a dominant dipolar contribution to the magnetic field topology.
Presently, the Herbig Ae star HD\,101412 possesses the strongest 
magnetic field ever measured in any Herbig Ae star, 
with a surface magnetic field $\left<B\right>$ up to 3.5\,kG. 
HD\,101412 is also the only Herbig Ae/Be star for which the rotational
Doppler effect was found to be small in comparison to the magnetic splitting 
and  several spectral lines
observed in unpolarised light at high dispersion are resolved into 
magnetically split components \citep{Hubrig2010}.

The finding that the dipole axis in HD\,101412 is located close to the stellar equatorial 
plane is very intriguing in view of the generally assumed magnetospheric accretion scenario that 
magnetic fields channel the accretion flows towards the stellar surface along the magnetic field lines.
As was shown by \citet{Romanova2003},
the topology of the channeled accretion critically depends on the 
tilt angle between the rotation and the magnetic axis. For large inclination angles $\beta$,
many polar field lines would thread the inner region of the disk, while the closed 
lines cross the path of the disk 
matter, causing strong magnetic braking, which could explain the observed unusually 
long rotation period of HD\,101412 of about 42 days.

Since about 70\% of the Herbig Ae/Be stars appear in binary/multiple systems
\citep{Baines2006}, special
care has to be taken in assigning the measured magnetic field to the particular component in the 
Herbig Ae/Be system.
\citet{Alecian2009} reported the discovery of a dipolar magnetic field in the Herbig~Be star
HD\,200775, which is a double-lined spectroscopic binary system. However, it should be noted that the 
magnetic field was discovered not in the component possessing a circumstellar disk and 
dominating the H$\alpha$ emission, so that the evolutionary status of the B3 primary component 
is yet unclear \citep{Benisty2013}. 
Similar to the case of HD\,200775, the frequently mentioned discovery of a 
magnetic field in the Herbig SB2 system 
HD\,72106  \citep{Alecian2009} refers to the detection only in 
the primary component, which is a young main-sequence star, but not in the Herbig Ae secondary 
\citep{Folsom2008}.
The same uncertainty in the evolutionary state applies to the magnetic field detection
in the system  V380\,Ori reported by \citet{Alecian2009}.
The authors detected the presence of a dipolar magnetic field of 
polar strength $2.12 \pm 0.15$\,kG on the surface of the chemically peculiar primary 
of the V380\,Ori system.
V380\,Ori has a spectral type around B9 and 
has been observed in great detail over many wavelength ranges
(e.g., \citealt{HamannPersson1992},
\citealt{Rossi1999}, \citealt{Stelzer2006}).
It has a close infrared companion, with
a separation of 0.15$^{\prime\prime}$ at PA 204$^{\circ}$ \citep{Leinert1997}. 
\citet{Alecian2009} found that the primary in the 
V380\,Ori system is itself a spectroscopic binary with
a period of 104\,days, with the secondary being a
massive T\,Tauri star. 
Most recently, \citet{Reipurth2013} reported that V380\,Ori is
a hier\-ar\-chi\-cal quadruple system with a fourth component at a distance of 8.8$^{\prime\prime}$ and 
position angle 120.4$^{\circ}$.
Since no periodicity was found in the behaviour of the emission in hydrogen, helium, calcium, 
and oxygen lines (the lines determining the Herbig~Ae/Be nature), it is possible that the primary 
chemically peculiar component with the detected 
dipolar magnetic field is already at an advanced age and that the Herbig Be 
status of the primary is merely 
based on the appearance of emission in the above mentioned lines belonging to the secondary T\,Tauri component.
 
Notably, the task of magnetic field measurements in Herbig stars is very challenging, 
as the work of \citet{Hubrig2015} demonstrates, in which the authors 
compiled all magnetic field measurements reported in 
previous spectropolarimetric studies. This study indicates that the low detection rate of magnetic 
fields in Herbig Ae stars, about 7\% \citep{Alecian2013}, can indeed be 
explained not only by the limited sensitivity of 
the published measurements, but also by the weakness of these fields. The obtained density distribution 
of the rms longitudinal magnetic field values reveals that only a few stars have magnetic fields stronger than 
200\,G, and half of the sample possesses magnetic fields of about 100\,G and less. 
Consequently, the currently largest spectropolarimetric survey of magnetic fields
in several tens of Herbig stars by \citet{Alecian2013}
using lower spectral resolution on ESPaDOnS and NARVAL cannot be considered as representative: 
the measurement accuracy in this study is worse than 200\,G for 35\% of the measurements, and 
for 32\% of the measurements it is between 100 and 200\,G.
Clearly, to improve our understanding of the origin of magnetic fields in 
Herbig Ae/Be stars and their interaction with the protoplanetary disk,
it is of utmost importance
to study magnetic fields with high accuracy measurements in a representative 
sample of Herbig Ae/Be stars.

\articlefigure[width=.7\textwidth]{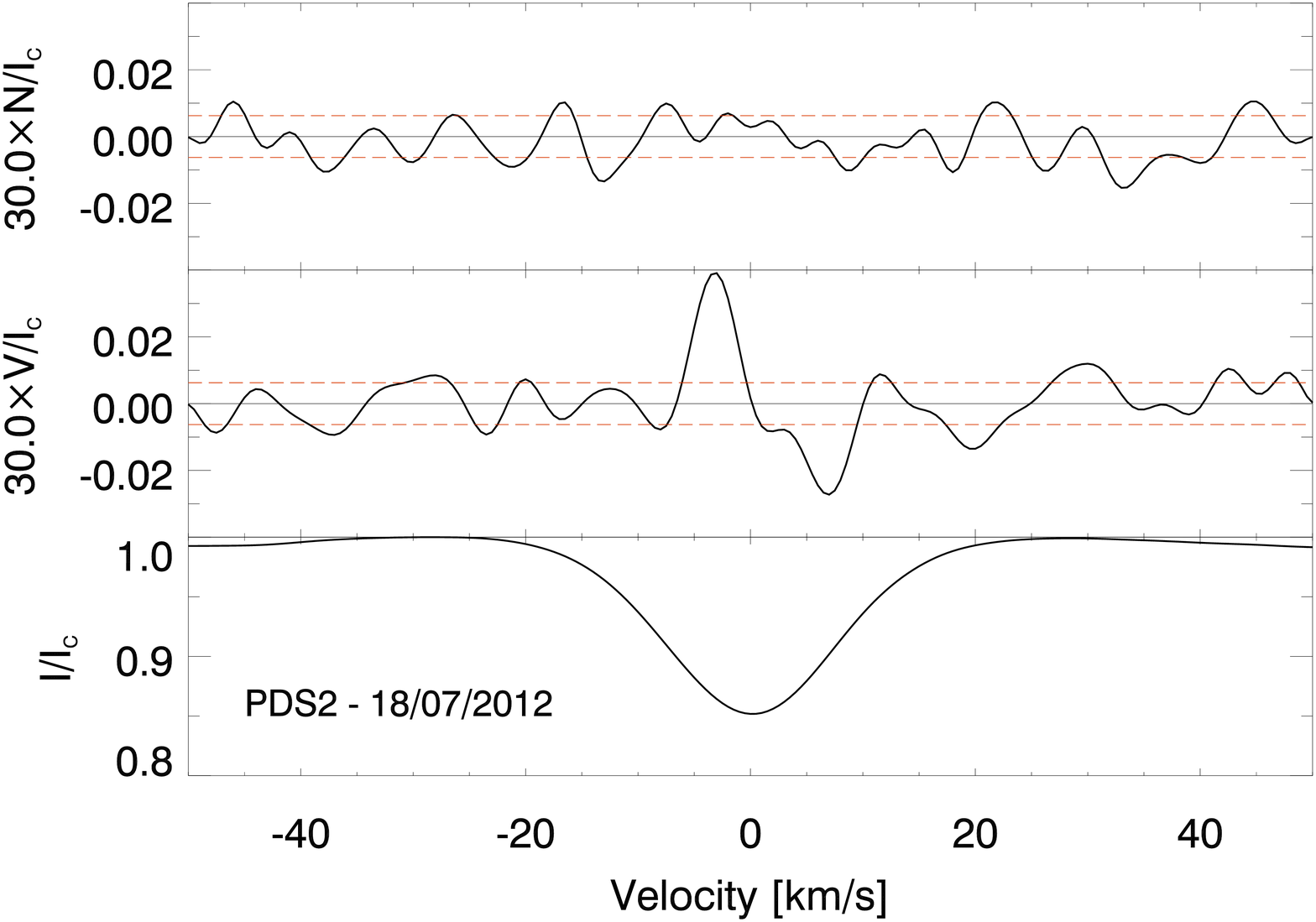}{fig:reduced}{Detection of a mean longitudinal 
magnetic field in the Herbig Ae star PDS\,2. 
The Stokes~$I$, Stokes~$V$, and diagnostic null ($N$) SVD profiles were used to determine
$\left<B_z\right>=33\pm5$\,G.
The Stokes~$V$ and $N$ profiles were expanded by 
a factor of 30 and shifted upwards for better visibility. The red 
dashed lines indicate the standard deviations for the Stokes~$V$ and $N$ spectra.
}

Zeeman signatures in the spectra of Herbig Ae/Be stars are generally very small, and increasing the 
signal-to-noise ratio (S/N) by increasing the exposure time is frequently limited by the shortness of the
rotation period of the star. Therefore, multi-line 
approaches as proposed by \citet{Semel1989} are commonly used to increase the S/N. The most widely used of these 
approaches is the Least Squares Deconvolution (LSD; \citealt{Donati1997}). The main assumption is the
application of the weak field approximation, that is, the magnetic splitting of spectral lines is 
assumed to be smaller than their Doppler broadening. Furthermore, it is assumed that the local line profiles 
are self-similar and can be combined into an average profile. Due to non-linear effects in the summation 
and the effect of blends, the resulting LSD profiles should not be considered as observed, but rather 
processed Zeeman signatures.
In a number of our studies we applied a novel magnetometry technique developed at the Leibniz-Institut
f\"ur Astrophysik Potsdam. The software package to study weak magnetic fields using 
the multi-line Singular Value Decomposition (SVD) method was introduced by \citet{Carroll2012}.
The basic idea of SVD is similar to the principal component analysis (PCA) approach, where the 
similarity of individual Stokes~$V$ profiles allows one to describe the most coherent and 
systematic features present in all spectral line profiles as a projection onto a small number 
of eigenprofiles. 
In Fig.~\ref{fig:reduced} we present an example of a very weak Zeeman feature in the HARPS\-pol
spectra of the Herbig Ae star PDS\,2 detected using the SVD method
\citep{Hubrig2015}.

\articlefigure{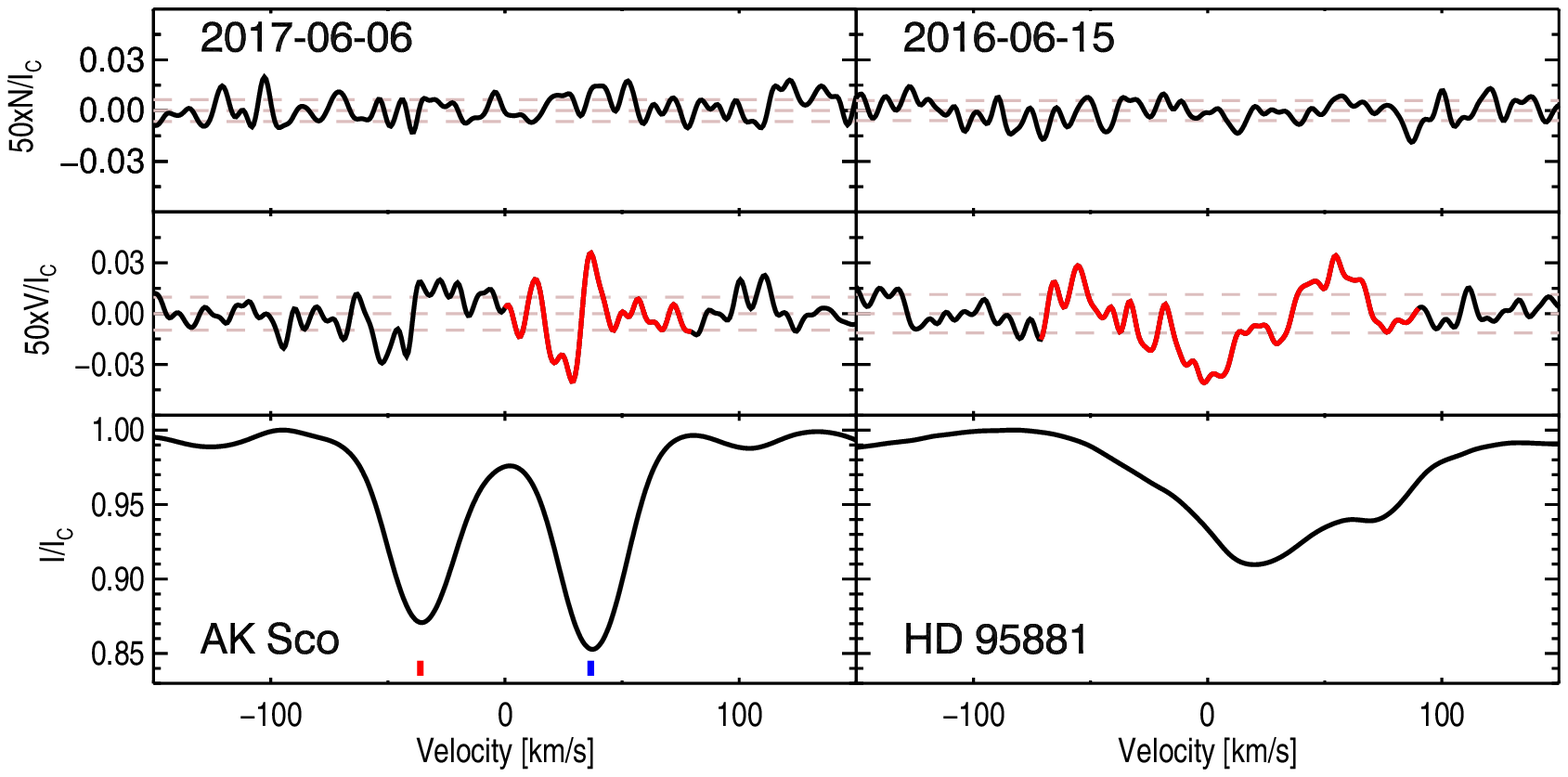}{fig:2}{SVD Stokes~$I$ (bottom), $V$ (middle), and diagnostic null ($N$) profiles (top).
The Stokes~$V$ and $N$ profiles have been amplified by a factor of 50. 
The detected Zeeman features are highlighted in red.
The horizontal dashed lines indicate the average values and the $\pm\sigma$--ranges.
{\emph Left:} AK\,Sco.
Under the Stokes~$I$ profiles, the components are marked with red (primary component) and blue (secondary component) ticks.
{\emph Right:} 
HD\,95881.}

In agreement with the merging scenario for the origin of magnetic fields in Ap stars, the number of close
binary systems with Ap components is very low:
only two such systems with Ap components, HD\,98088 and HD\,161701, are known 
\citep{Abt1968,Hubrig2014}. Similarly,  studies of Herbig Ae stars
by \citet{Wheelwright2010} and \citet{Duchene2015}
indicate the lack of close binary systems 
with P$_{\rm orb}<20$\,d. Weak magnetic fields were detected in two Herbig Ae systems, in AK\,Sco and in the presumed binary 
HD\,95881, using HARPS\-pol \citep{Jarvinen2018}. Using high quality HARPS\-pol observations, we
obtained $\left<B_z\right>=-83\pm31$\,G for the secondary component in the system AK\,Sco and 
$\left<B_z\right>=-93\pm25$\,G for HD\,95881. Examples of their Stokes~$I$, Stokes~$V$ and
diagnostic $N$ profiles are presented in Fig.~\ref{fig:2}. 
It is of interest that for AK\,Sco we observe the magnetic field in the secondary component
in the region of the stellar surface facing permanently the primary component, meaning that the 
magnetic field geometry in the secondary component is likely related to the position of the 
primary component. We note that a similar magnetic field behavior, where the field orientation is 
linked to the companion, was previously detected in HD\,98088 and HD\,161701, the
two close main-sequence binaries with Ap components mentioned above. 
Further, our recent  re-analysis of 
HARPS\-pol observations of another SB2 system, HD\,104237 confirmed that  both components, the Herbig Ae star
and the lower mass T\,Tauri star possess magnetic fields (J\"arvinen et al., {\it in preparation}).
Obviously, a search for magnetic fields and the determination of their geometries in close binary systems  
is very important as the knowledge of the presence of a magnetic field and of the
alignment of the magnetic axis with respect to the orbital radius vector in Herbig binaries 
may hint at the mechanism of the magnetic field generation.

\section{Magnetospheric accretion and the link with observed spectral properties}

The weakness of the observed magnetic fields put into question our current understanding of the 
magnetospheric accretion process in intermediate-mass pre-main sequence stars.
Importantly, \citet{CauleyJohnsKrull2014} 
studied the He~{\sc i}~$\lambda$10830 morphology in a sample of 
56~Herbig~Ae/Be stars. They suggest that early Herbig~Be stars do not accrete material from their inner
disks in the same manner as T\,Tauri stars, while late Herbig~Be and Herbig~Ae stars show evidence for 
magnetospheric accretion. Furthermore, they proposed more compact magnetospheres in Herbig Ae/Be 
stars compared to T\,Tauri stars.

Most recently, \citet{Ababakr2017} presented  
H$\alpha$ linear spectropolarimetry of a sample of 56~Herbig Ae/Be stars.
A change in linear polarization across this line was detected in 42 (75\%) objects, indicating
that the circumstellar environment around these stars on small spatial scales has an asymmetric 
structure, which is typically identified with a disk. A second outcome of their research was 
the confirmation that Herbig Ae stars are similar to T\,Tauri stars in displaying a line polarization 
effect, while depolarization is more common among Herbig Be stars. 

Using near-infrared multi-epoch spectroscopic data obtained with the CRIRES and X-shooter spectrographs 
installed at the VLT,
\citet{Schoeller2016} examined the magnetospheric accretion in the Herbig~Ae star HD\,101412.
Spectroscopic signatures in He\,{\sc i}~10830 and Pa$\gamma$,
two near-infrared lines that are formed in a Herbig star's accretion region, were presenting temporal modulation.
The authors showed that this modulation is governed by the rotation period of this star and the observed
spectroscopic variability was explained within the magnetic geometry
established earlier from magnetic field measurements by \citet{Hubrig2011}.

\articlefigure[width=.7\textwidth]{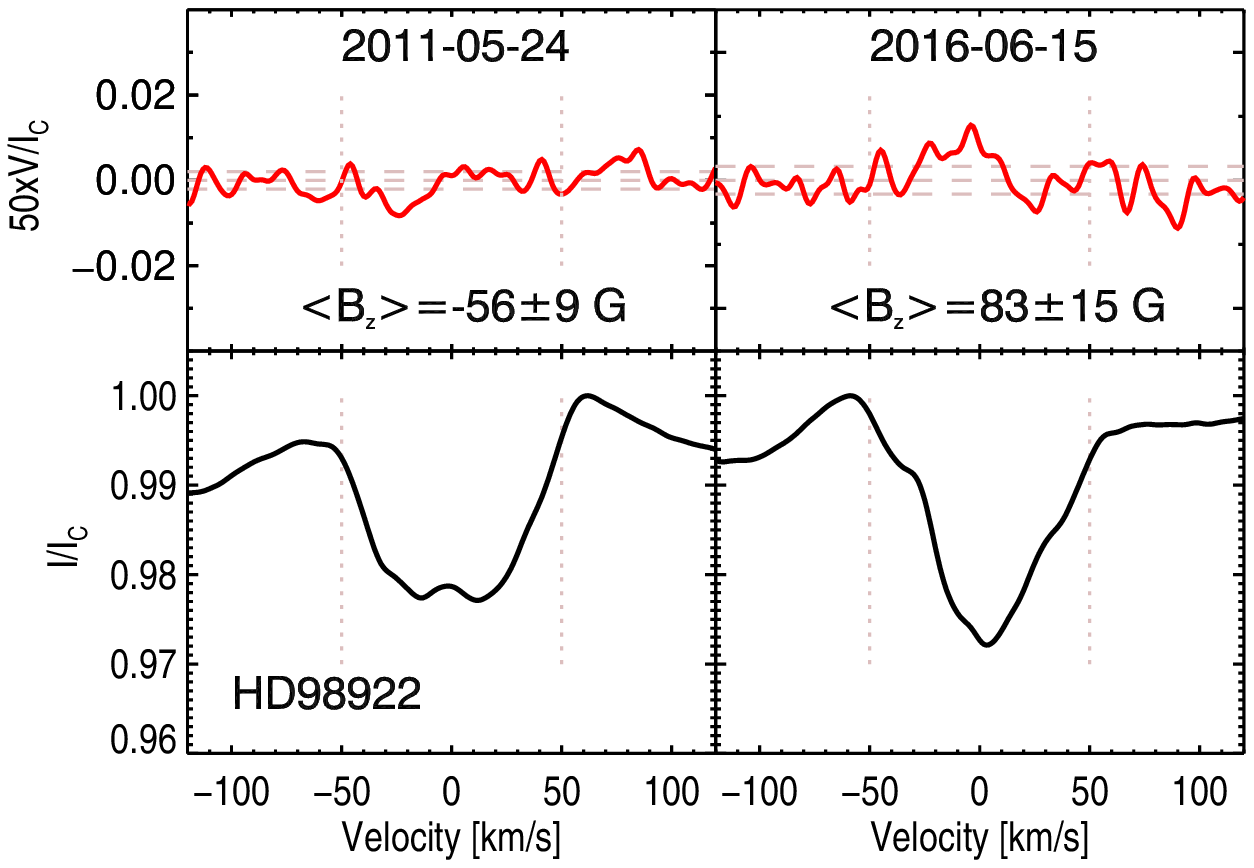}{fig:3}{LSD Stokes~$I$ (bottom) and Stokes~$V$ (top) profiles
obtained 
for HD\,98922 on two different observing epochs. The Stokes~$V$ profiles have been amplified by a factor of 50. 
The horizontal dashed 
lines indicate the average values and the $\pm\sigma$--ranges.}

An interferometric search for magnetospheres by
\citet{Kraus2008} made use of the observed Br$\gamma$ line emission in five Herbig Ae/Be stars with 
AMBER on the Very Large Telescope Interferometer
and found only in HD\,98922 (then not known to be magnetic) a small size compatible 
with a magnetosphere.
For the other four sources (including the magnetic SB2 system HD\,104237) they found larger sizes 
consistent with an extended stellar wind or a disk-wind. As already mentioned above, both components
in the system HD\,104237 possess magnetic fields, and it is not clear why the interferometric 
observations failed to detect their magnetospheres. Our recent HARPS\-pol observations of HD\,98922 presented in 
Fig.~\ref{fig:3} confirmed the presence of a weak magnetic field.
The high $S/N$ observations presented in this figure clearly indicate that
the intensity profiles in Herbig stars are 
highly variable. The strong spectral variability of these stars is  due to the complex interaction between 
the stellar magnetic field, 
the accretion disk, and the stellar wind. Therefore, any search for the presence of magnetic fields 
should involve multi-epoch spectropolarimetric observations. Given the small number of visits
in the study of \citet{Alecian2013}, several magnetic fields were certainly missed. This is illustrated
in their study of HD\,35929, which only on one occasion out of five shows a magnetic field. Thus, it is likely
that a larger number of the supposed non-magnetic stars are also magnetic.

Ignoring the presence of the spectral and magnetic variability observed in  Herbig stars,
\citet{Reiter2018} used single-epoch He\,{\sc i} $\lambda$10830 line profiles
for 64~Herbig stars with magnetic field measurements mentioned in the work of  \citet{Alecian2013} in
order to test the presence of magnetospheric accretion.
However, unless we directly look onto the rotation pole, or the obliquity angle is zero, we expect
rotational modulation of the magnetic field and the spectral line profiles. It was also not clear in this study
to which aspect of the (unknown) magnetic field geometry these spectra belong. 
Not surprisingly, the authors did not find any correlation of the line profile morphology
with the luminosity, rotation rate, mass accretion rate, or disk inclination. 
Clearly, we will understand the magnetic field and its interaction with the stellar environment only
with good coverage of the rotation cycle.

\section{Future prospects in studies of Herbig Ae/Be stars}

It appears that the as yet small number of magnetic Herbig Ae/Be stars can be due to the weakness of 
the magnetic fields and/or the large measurement errors. 
According to \citet{Alecian2014}, the
magnetic properties of A and B-type stars must have been shaped before the Herbig Ae/Be phase of the stellar evolution.
Using pre-main-sequence evolutionary tracks calculated with the CESAM code  \citep{Morel1997},
she concluded that even stars above three solar masses will undergo a purely convective phase before 
reaching the birthline.
Therefore, it is reasonable to assume that the weak magnetic fields detected in a number of Herbig Ae/Be stars are
just leftovers of the fields generated by pre-main-sequence dynamos during the convective phase.
If this scenario is valid, we should expect a significantly larger number of Herbig stars possessing weak 
magnetic fields.

Spectropolarimetric observations of a sample of 21 Herbig Ae/Be stars observed with FORS\,1
have been used to search for a link between the presence of a magnetic field and other stellar
properties \citep{Hubrig2009}.
This study did not indicate any correlation of the strength of the longitudinal magnetic 
field with disk orientation, disk geometry, or the presence of a companion. 
No simple dependence on the mass-accretion rate was found, but the range of the observed field values 
qualitatively supported the expectations from magnetospheric accretion models with dipole-like 
field geometries. Both the magnetic field strength and the X-ray emission showed hints of 
a decline with age in the range of $\sim2-14$\,Myr probed by the sample, supporting a dynamo 
mechanism that decays with age. Importantly, the stars seemed to obey the universal power-law 
relation between magnetic flux and X-ray luminosity established for the Sun and main-sequence 
active dwarf stars \citep{Pevtsov2003}. Future work on stellar properties of magnetic Herbig stars should involve 
a larger and more representative sample to determine the existing relations at a higher
confidence level.

While magnetic surveys of Herbig Ae/Be stars so far have mainly targeted
the detection of the magnetic fields, we should try to understand their three-dimensional
structure with respect to the stellar rotation axis and the disk orientation. Importantly, such a structure
can only be studied by monitoring the targets over a significant number of nights to sample their
rotation periods. 
Spectropolarimetric observations of photospheric lines
are usually used to determine the geometry of the global magnetic field, while spectropolarimetry 
of accretion diagnostic lines (e.g., He~{\sc i} 5876\,\AA{}, the Na\,{\sc i} doublet, and the Balmer lines) probes the 
accretion topology in the accretion columns, which are perturbed by interactions with the disk. 
From the temporal variations
of the measured longitudinal magnetic fields, making a frequency analysis and computing best-fit curves 
to these variations, we will be able to ascertain their rotation/magnetic periods and compare
them with those obtained from fundamental parameters (e.g.\ \citealt{Hubrig2009}).
The Doppler-shifted spectropolarimetric contributions
from photospheric and circumstellar environmental diagnostic lines will allow to apply the 
technique of Zeeman-Doppler tomography to determine the correspondence between the magnetic field 
structure and the radial density and temperature profiles. 

From the resulting knowledge of the magnetic fields and the position of the Herbig Ae/Be
stars in the H-R diagram, it will become possible in the future to study the evolution of the
magnetic field topology and its stability across the PMS tracks, the impact of magnetic fields 
on the evolution of the rotation rates, and possible correlations between
evolutionary state and other stellar properties. 
The study of the magnetic field topology and the evolution of the magnetic field structure on the 
surface of Herbig Ae/Be stars will furthermore
lead to a deep insight into the complex interaction between the stellar magnetic field, 
the accretion disk, and the stellar wind. It will also provide crucial additional information to 
test the predictions of existing theories on the origin of the magnetic field and its role in 
the star and planet formation process.

\acknowledgements MAP acknowledges the support of the Foundation "2017-RFBR-AZ-2" (grant number 18-52-06004).

\end{document}